\documentstyle[aps,prl,epsfig]{revtex}

\def\a{{\alpha}}
\def\L{{\Lambda_{UV}}}
\def\l{{\lambda}}
\def\ba{\begin{eqnarray}}
\def\ea{\end{eqnarray}}
\def\nnb{\nonumber}
\draft

\begin{document}
\twocolumn[\hsize\textwidth\columnwidth\hsize\csname
@twocolumnfalse\endcsname

\title{Radiative electroweak symmetry breaking
in the extra dimensions scenarios}

\author{Qi-Shu Yan}

\address{Physics Department of Tsinghua University, Peoples' Republic of China}

\date{\today}

\maketitle

\begin{abstract}
We study the radiative spontaneous electroweak
symmetry breaking in the extra dimensions
scenarios of the standard model extension proposed
by Antoniadis {\it et al}, Dienes {\it et al.} and
Pomarol {\it et al.}. In the framework of
multi-scale effective theory when viewing from
the ultraviolet cutoff scale down to the low
energy scale, we find that
the effects of Kaluza-Klein excitations of bosons can
change the sign of
the Higgs mass term of the standard model from
positive to negative and therefore trigger
the electroweak symmetry breaking at 1.6 (2) TeV
or so if the compactification scale is assumed to
be 0.8 (1.5) TeV or so. New particle contents beyond
the SM or supersymmetry are not necessary for this
mechanism. We conclude that in the extra dimension
scenarios, the radiative correction can naturally
induce the desired electroweak symmetry breaking.
\end{abstract}

\vskip1.9pc]


The extra dimension (ED) scenario of the
standard model (SM) extension is fascinating
and is under the vigorous
investigation (refer \cite{Nath:2000kr} for a brief
review), since it provides
a new paradigm which is within
the reach of near future experiments
and beyond technicolor and
supersymmetry (SUSY). Many efforts have
been invested in the ED model
construction \cite{Kawamura:2000ir}.
However, a realistic model
requires to answer how to break
the electroweak symmetry.

In the SM, to trigger the spontaneous
electroweak symmetry breaking (SEWSB),
the mass term of the Higgs potential
is set to be negative by hand.
Although the Higgs
mechanism is sucessful in explaining the
masses of bosons and fermions and their mixing,
it is quite unnatural in this respect.
The underlying reason for the SEWSB of the SM
is still an open question \cite{Chanowitz:1988ae}.

There are two main models to explain the SEWSB
of the SM: technicolor and SUSY
models. In technicolor models, Higgs field is
regarded as a composite field of other more
fundamental fields. The SEWSB is due to the
fermion condensate. The SEWSB of the
technicolor version in the ED has been
intensively investigated (refer \cite{cheng}
for a review).

In SUSY models, the radiative symmetry breaking
mechanism \cite{Coleman} is
well known \cite{Lahanas:1987uc}.
Due to the largeness of
top Yukawa couplings, the
mass term of the Higgs doublet coupled
to u-type quarks can be driven from positive to
negative and the SEWSB
can be triggered naturally. However,
we would like to point out that beside the
the largeness of top Yukawa coupling, another
important factor which is always under-emphasized
is the fact that in softly broken SUSY models there
exist couplings with mass dimension
(trilinear couplings, for instance) and
masses of heavy superpartners (especially the
scalars of top-quark partner which contribute
to the radiative SEWSB constructively).
Without these important terms, the radiative
breaking mechanism of the SUSY can not work.

In some ED scenarios, there are a lot of heavy
particle spectrum in the reduced 4D effective theory
(For example, the Kaluza-Klein (KK) excitations of both vector
and Higgs boson are possibly quite heavy). Is it
possible for these heavy
particles to induce the desired SEWSB at a few TeV,
as the scalars of SUSY do? In this work, we will
answer this question and study the radiative
mechanism in non-supersymmetric ED extension of
the SM proposed by
Antoniadis {\it et al} \cite{Antoniadis},
Dienes {\it et al.} \cite{Dienes:1999vg} and
Pomarol {\it et al.} \cite{Pomarol:1998sd}, where
gauge and Higgs bosons are assumed to propagate
in the bulk. The case where only vector bosons
live in the bulk and and the universal case
where all fermions are also assumed to live
in the bulk are discussed

Firstly, we examine how it
is possible to spontaneously break symmetry in
a two-real-scalar system.
And we use the multi-mass-scale effective potential
method (MEPM) given in the refrences
\cite{bando:1993} and \cite{Casas:1999cf}. In
a theory with
more than one scale, MEPM can avoid large logarithms
and preserves the validity of perturbation theory.
The basic ingredients of the MEPM include
the renormalization group equation (RGE) method
and the decoupling
theorem \cite{appelquist}.
One of the advantages of the MEPM is that
by using different effective field
theories \cite{georgi} which are
dependent on the field scales,
it is convenient to solve the RGEs
of the effective potential. Because
the RGEs are just the same ones in each
interval between mass thresholds.

The effective Lagrangian of
the system defined at the ultraviolet cutoff
scale (UV) is assumed to be
\ba
L_{UV}&=&\frac{1}{2} [\sum_{i=1}^2 (\partial^{\nu} H_i)^{\dagger} \partial_{\nu} H_i - m_{H_i}^2(\mu_{UV}) H_i^{\dagger} H_i]\\\nnb
&&-V(H_1,H_2)\,,
\ea
where $H_i, i=1,2$ are two real scalar fields.
We assume $|m_{H_1}^2(\mu_{UV})|< m_{H_2}^2(\mu_{UV})$.

Motivated by the effective 4D Lagrangian reduced
from the 5D one, the potential $V(H_1, H_2)$ is
simply assumed to have the form
\ba
V(H_1,H_2)&=&\frac{\l(\mu_{UV})}{4!} (H_1^4 + 6 H_1^2 H_2^2 + H_2^4)\,,
\ea
where $\l(\mu_{UV})$ is positive.
The potential is invariant under the transformation
$H_1->- H_1$ and $H_2->- H_2$.
In the UV, We can choose $m_{H_i}^2(\mu_{UV}), i=1,~2$,
and $\l(\mu_{UV})$ as the three free parameters
owned by the system. The symmetry is assumed to
be unbroken and
$m_{H_i}^2(\mu_{UV}) $ are positive.

At the low energy scale (IR),
after $H_2$ decouples at its threshold scale
$m_{H_2}$\footnote{Although $H_2$ should decouples at
$M_{H_2}=m_{H_2}+\l H_1/2$ as stated in
\cite{Casas:1999cf}, due to the large mass of
KK excitations, the difference between $m_{H_2}$
and $M_{H_2}$ is omitted here.}, the obtained
effective theory at low energy scale $\mu_{IR}$ can
be written as
\ba
L_{eff}&=&\frac{1}{2}(\partial^{\nu} H_1)^{\dagger} \partial_{\nu} H_1 - m_{H_1}^2(\mu_{IR}) H_1^{\dagger} H_1 \\ \nnb
&&-V(H_1) + ...\,,
\ea
where the dots represent the omitted irrelevant terms
and $V(H_1)$ can be simply expressed as
\ba
V(H_1)&=&\frac{\l_1(\mu_{IR})}{4!} H_1^4\,.
\ea
This effective Lagrangian also owns a $Z_2$
reflection symmetry.

The one-loop RGEs of this system are listed below
\ba
\frac{d m_{H_1}^2(t)}{d t } &=& \frac{1}{4 \pi} [\a_\l m_{H_1}^2(t)
+ \a_\l m_{H_2}^2(t) \theta(\mu - m_{H_2})]\,,\\
\frac{d m_{H_2}^2(t)}{d t } &=& \frac{1}{4 \pi} [\a_\l m_{H_1}^2(t)
+ \a_\l m_{H_2}^2(t)] \theta(\mu - m_{H_2})\,,\\
\frac{d \a_\l}{d t} &=& \frac{1}{4 \pi} [3 \a_\l + 3 \a_\l \theta(\mu - m_{H_2})]\,,
\ea
where $t=ln \mu/\mu_{IR}$, $\a_\l=\l/(4 \pi)$,
and $\theta(\mu-m_{H_2})$ is
the Heaviside theta function.

At the IR, we
assume that $m_{H_1}^2(\mu_{IR})$ is negative
and the $Z_2$ symmetry is broken.
The IR free parameters
can be chosen as $v$ (i.e. the vacuum expectation
value (VEV) of the field $H_1$), $\a_\l(\mu_{IR})$,
and $m_{H_2}$. The $\a_\l(\mu_{IR})$ should
be positive as required
by the stability of the vacuum.

The value of $m_{H_1}(\mu_{IR})$ is fixed by the
relation $m_{H_1}^2(\mu_{IR})=- 2 \pi/3 \a_\l(\mu_{IR}) v^2$.
By solving the RGEs given in eqs. (5-7),
it is obvious that below the threshold value of
$m_{H_2}$, the self coupling $\l$
will become stronger and stronger with the running of
energy scale and the negative mass term $m_{H_1}^2$
will be simply driven to be further negative.

When the RGEs run across the threshold
value of the $m_{H_2}$, the degree of freedom
of the $H_2$ is activated. If the condition
$|m_{H_1}^2|(t)|<|m_{H_2}^2(t)|$ is
satisfied for $\mu \geq m_{H_2}$,
it is sufficient for the existence of
a scale $\mu_{cri}$ where the sign
of $m_{H_1}^2$ can be flipped from negative to
positive. At the scale $\mu_{cri}$, the
VEV of the $H_1$ become zero and
the broken symmetry is restored.
This condition can be solved out from
eqs. (5-7) as
\ba
|\frac{2 \pi}{3} \a_\l v^2| < m_{H_2}^2 (1-\frac{\a_\l}{4 \pi} \ln\frac{m_{H_2}^2}{\mu_{IR}})^{1 \over 3}\,.
\label{cond}
\ea

Therefore, in the IR, for a fixed $v$,
choosing a proper point in the parameters space of
$m_{H_2}$ and $\a_\l$ which satisfies the
condition given in eqn. (\ref{cond}) as an IR input,
and running the RGEs from the IR up to the UV,
we can get a point of in the parameter space
determined ( $m_{H_i}^2(\mu_{UV}), i=1,~2$ and
$\a_\l(\mu_{UV})$) in the UV.
With this set of parameters
as an UV input, and running the RGEs from the UV
down to the IR, we can definitely get the
desired radiative spontaneous symmetry breaking
at the critical scale $\mu_{cri}$.

The lesson we learn from this case is that
if an IR input of a system can trigger the symmetry
restoring, then when running from up down to bottom,
the UV input determined by the IR input
can definitely trigger the symmetry
breaking at the same critical scale, since
the RGEs are differential equations and
are solvable with a specified boundary condition.

Now let's examine the
case in the ED scenarios.
We consider a simple extension of the SM to
5D \cite{Pomarol:1998sd} (It is straightforward to
generalize our discuss to cases with high number
of extra dimensions). The fifth space-like
dimension $x_5$ is assumed to compactify on
the orbifold $S^1/Z_2$. The
5D Lagrangian is defined as
\begin{eqnarray}
{\cal L}_{5D}&=&-\frac{1}{4} F_{MN}^2+ |D_M H|^2-V(H)\nonumber\\
&&+L_{GF}+\Big[i \bar \psi_{i}\sigma^\mu D_\mu \psi_{i}+ Y_F \bar \psi_{L} H \psi_{R} \Big ] \delta(x_5)\,,
\label{lagrangian}
\end{eqnarray}
where $F_{MN}=\partial_{M} A_{N}-\partial_N A_{M}
+ g_5 [A_{M}, A_{N}]$, which is the gauge field tensor
defined in 5D, and
$N, M = 0, 1, 2, 3, 5$. The group
generators index is omitted.
Gauge fields $A_{M}$ and Higgs weak doublet
field $H$ have mass dimension $3/2$.
$g_5$ and $Y_{F}$
are the gauge coupling and
the Yukawa coupling, respectively,
and have mass dimension $-1/2$.
${\cal L}_{GF}$ is the gauge fixed term.
$V(H)$ is the usual Higgs potential and
has the form
\begin{eqnarray}
V(H)&=&\mu^2 H^\dagger H+{\l \over 4}(H^\dagger H)^2\, ,
\end{eqnarray}
$\mu^2$ and $\l$ are the mass term
and self coupling of H boson, respectively. And
$\l$ has mass dimension $-1$. Here $\mu^2$ is
assumed to be positive and the
$SU(2) \times U(1)$ symmetry is assumed to be
preserve when the compactification occurs.
According to the power counting law,
the gauge theory defined in 5D is
non-renormalizable, due
to the fact that the couplings have
negative mass dimension.

The fields living in the
bulk can be defined to be even
under the $Z_2$-parity and can be Fourier-expanded as
\begin{eqnarray}
A_{M}(H)(x_\mu,x_5)&=&\sum^{\infty}_{n=0}
\cos\frac{nx_5}{R_c}A_{M(5D)}^n(H_{(5D)}^{n})(x_\mu)\, ,
\label{fourier}
\end{eqnarray}
where $R_c$ is the compactification size of the fifth
dimension, and $A_{M(5D)}^{(n)}(H_{(5D)}^{n}), n\neq 0$ are
KK excitations. Zero modes are
localized on the 3-brane and are fields
defined in the SM. Substituting
eqn. (\ref{fourier}) into eqn. (\ref{lagrangian}),
and rescaling fields and parameters with
$\l_{5D} = 2 \pi R \l_{4D}$,
$g_{5D}(Y_{u5D})=\sqrt{2\pi R}\, g_{4D}(Y_{u4D})$,
$A_{\mu5D}^0(H_{5D}^{0})=A_{\mu4D}^0(H_{4D}^{0})/\sqrt{2\pi R}$,
$A_{\mu5D}^{n}(H_{5D}^n, A_{55D}^n)=A_{\mu4D}^{n}(H_{4D}^n, A_{54D}^n)/{\sqrt{\pi R}}\: ,\; (n\not =0)$.
Then we get the effective Lagrangian in 4D
defined at the UV $\L$ which has a form
\begin{eqnarray}
{\cal L}^{eff}&=&{\cal L}_{SM} + \delta {\cal L}^{ED}\,,
\end{eqnarray}
where ${\cal L}_{SM}$ is just the Lagrangian
of the SM in 4D,
and $\delta {\cal L}^{ED}$ contains all
interactions of KK excitations with zero
modes on 3-brane. The effective theory
owns a $SU_c(3)\times SU_L(2)\times U_Y(1)$
symmetry. In order to
be compatible with the SM,
$A_5$ (the fifth component of vector boson field)
is assumed to have no zero mode.
So by utilizing the standard dimension
reduction procedure, we specify a
basic effective Lagrangian of extra
dimension scenarios with the symmetry of
the SM.

Two features of the model are
remarkable \cite{Dienes:1999vg}. The first one
is universal KK excitation spectrum of
fields propagating in the bulk. The
mass values of each level of KK
excitations can be expressed as
$m_i=i/R_c$,
and are independent of other
quantum numbers, spin and charge,
for instance.

The second one is that the presence of
infinite towers of KK states makes
the effective theory
non-renormalizable and this is an
intrinsic feature for high dimensions theory.
In order to have a renormalizable
effective theory, an explicit ultraviolet cutoff
$\Lambda$ is introduced to truncate the
infinite KK excitation.

The RGEs of gauge couplings take the form
${d \a_g}/{d t}=2 b_g \a_g/(2 \pi) $,
and up to one-loop level, $b_g$ is defined as
\ba
b_g&=&- \frac{11}{3} C_g(G) + \frac{2}{3} \sum_{f} T_g(\Psi_f) + \frac{1}{3} T_g(H)\nnb\\
&&+[- \frac{11}{3} C_g(G) + \frac{1}{6} C_g(A_5) + \frac{1}{3} T_g(H)] N_{KK}\,,
\ea
where $\a_g=g^2/(4 \pi)$, and $N_{KK}= \sum_{i=1} \theta(\mu - m_i)$,
counting the number of activated KK excitations.
To understand the RGEs of gauge couplings,
it is noticeable that the excitations KK states
of vector fields $A_{\mu}$ and scalar fields $A_5$
are the adjoint representations of gauge groups,
while the complex weak doublet $H^{(n)}$ are
the fundamental representations of $SU(2)$.

It is remarkable that, in the extra dimension
scenarios we consider here, KK excitations
always drive the 4D gauge couplings of
$SU(3)$ and $SU(2)$ groups to their
weak coupling limits.
This is just the typical feature of
non-Abelian gauge theory, i.e. the asymptotical
freedom. Even for the case that
all fermions in the SM (the universal case)
are assumed to live in the bulk, the couplings of
$SU(3)\times SU(2)$ are driven to their weak
coupling limit.

To simplify our analysis,
below we will omit the contributions
of the $U(1)$ group,
due to its small effect to the problem we
consider here.

The Yukawa coupling terms of quarks in
the SM are assumed to be the contact terms
and have the form
\ba
L_{QUH}&=&Y_t {\bar Q_L} H U_R +h.c.\,,
\ea
here ${\bar Q_L}=({\bar u_L}, ~{\bar d_L})$,
$H^T=(H^0,~H^-)$, and $Y_F$ is the Yukawa couplings.

Due to its large effects, the
Yukawa coupling of top quarks should be
considered.
The RGE of it can be written as
$d \a_h /d t = b_h \a_h /(2 \pi)$, and up to one-loop
level $b_h$ is
defined as
\ba
b_h &=& - 6 C_c (1+2 N_{KK}) \a_3 - 3 C_w (1+ 2 N_{KK}) \a_2 \nnb\\
&&+[N_c + \frac{3}{2} (1+ 2 N_{KK})] \a_h \,,
\label{betah}
\ea
where $\a_h=Y_t^2/(4 \pi)$, and $N_c$ is
the number of color.
The $2$ before $N_{KK}$ is due to
the different normalization
of zero modes and KK excitations.
$C_c$ and $C_w$ are quadratic Casmir
operators of $SU(3)$ and
$SU(2)$ groups for fundamental
representations, respectively.

The RGE of self-interaction
coupling of Higgs zero mode fields $\l$
can be expressed as $d \a_\l/d t =b_{\l} \a_\l /(2 \pi)$,
and up to one-loop, $b_{\l}$ is defined as
\ba
b_{\l} &=&3 (1+N_{KK}) \a_{\l} + \left [(2 C_w +4 )(1+ N_{KK}) \right .\nnb \\
&&\left. + \left( 4 C_w^2 + (2 D_B + 8) c^{\prime} \right ) \frac{\a_2}{\a_{\l}} (1+N_{KK}) \right ] \a_2\nnb\\
&&- 2 N_c (2 \frac{\a_h}{\a_{\l}} - 1) \a_h\,,
\label{betal}
\ea
where $D_B=5$, $\a_{\l}=\l/(4 \pi)$, and
$c^{\prime}=1 + N_w - 2/N_w + 1/N_w^2$.
$D_B$ is to count the degree of freedom
of the gauge vector bosons.
For $SU(2)$, $N_w=2$. The first two terms in Eqn.
(\ref{betal}) tend to increase $\a_l$,
and the last term tends to decrease it.
While KK excitations of bosons always
tend to drive $\a_\l$ to be large.

The one-loop RGE of Higgs mass term is expressed
in the below form
\ba
\frac{d m_H^2}{d t}&=& \frac{1}{2 \pi} \left \{ (\a_\l-3 C_w \a_2 + N_c \a_h) m_H^2\right. \nnb\\
&&\left .+(\a_\l -3 C_w \a_2 ) N_{KK} m_H^2 \right. \nnb\\
&&\left .+\left((\a_\l+ (D_B - 2) C_w \a_2\right) M_{KK}^2 \right \}\,,
\label{mssrge}
\ea
where $M_{KK}^2=\sum_{i=1} m_i^2 \theta(\mu-m_i)$,
counting the contributions of KK excitations
to the renormalization constant of $m_H^2$.

In the UV, there are six free parameters in the
effective theory needed to be specified,
1)$\a_3$ and 2) $\a_2$, the fine structure
constants of gauge group $SU(3) \times SU(2)$,
3) $\a_h$, 4) $\a_l$ and 5) $m_{H}^2$ and 6) $M_c$.
While in the low energy scale, there are
only two free parameters needed to be specified
1) $\a_l$, and 2) $M_c$, since
$\a_3$, $\a_2$, $\a_h$, and  $m_{H}^2$
can be determined from experiments or
those three free parameters.
The $M_c=1/R_c$ determines that where
KK excitations should be counted.
While triviality and stability conditions of Higgs
potential in the SM constrain the value
of $\a_\l(\mu_{IR})$ \cite{hambye}.
The Landau pole of $\a_\l$ can fix
the value of $\L$.

We will concentrate on the analysis of
the behavior of $m_H^2$ from the IR up to
the UV. At the low energy scale (say
$\mu_{IR}=M_z$, the $m_H^2$ is chosen to be
negative as required by the SM.
And its initial value at IR can be fixed by
$v$ (the VEV of $H$) and the free
parameter $\a_\l(\mu_{IR})$.

Before the running scale cross the
threshold of the first
KK excitations, only the first
term in Eqn. \ref{mssrge} contributes which tends
to drive $m_H^2$ to be further negative. After
only few KK excitations are counted,
$m_H^2$ can change its sign from negative to
positive, as shown in
figure 1.

\begin{figure}[h]
\begin{center}
\epsfysize=7truecm\epsfbox{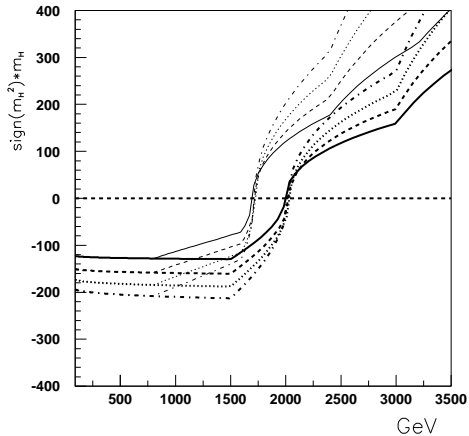}
\vspace{0truecm}
\end{center}
\caption{The varying of the sign of $m_H^2$ and the
value of $m_H$ with the energy scale. The sold, dashed
dot, and dash-dot lines represent $\l_(M_z)=1.0$,
$\l_(M_z)=1.5$, $\l_(M_z)=2.0$ and $\l_(M_z)=2.5$,
respectively. The group of wide(thin)
lines corresponds to the case $M_c=0.8$ TeV ($M_c=1.5$ TeV).}
\label{fig1}
\end{figure}

As we know, in the SM, the sign of the mass
term $m_H^2$ completely
determines whether the electroweak symmetry
is broken or unbroken. The change of $m_H^2$
from the negative to the positive value
 hints the restoring of the broken symmetry.
With the same reasoning, as shown in our toy model,
we can conclude it is possible to have the radiative
SEWSB mechanism in the ED scenarios
of the SM.

The underlying reason for the radiative
SEWSB is that the KK
excitation of bosons are very heavy.
With an appropariate value of
$\a_\l(\mu_{IR})$, the process of the
symmetry restoring can quickly
happen after the running scale has crossed
the threshold of the first KK excitation.

We also check the case in
which only vector bosons live in the bulk, and
find that with properly chosen $M_c$ and $\l$,
the radiative breaking mechanism exists too.
For the universal case,
we find that due to the large contribution from
KK excitations of top KK excitations
and the large top Yukawa coupling,
it is relatively hard to find a approperate point
which can both invoke SEWB and
preserve the validity of perturbation.
However, in the universal case of the two
Higgs doublets extension of the SM, with the help
of $\tan\beta$ (the ratio of the VEV of the two
Higgs doublets), it is still possible to have
the radiative SEWSB mechanism induced by KK bosons.

In summary, we investigate the radiative
electroweak symmetry breaking in the
extra dimensions scenarios of the
SM extension proposed by
Antoniadis {\it et al.},
Dienes {\it et al.} and Pomarol {\it et al.}.
Utilizing the decoupling theorem and
the one-loop renormalization group equations
of the parameters
of Higgs potential, we find
that heavy KK bosons
can change the sign of the Higgs mass terms
from positive to negative and therefore trigger
the SEWSB. We conclude that the radiative
mechanism can naturally exist in the
ED scenarios, and new particle contents
beyond the SM or SUSY are not necessary.


\begin{references}

\bibitem{Nath:2000kr}
P.~Nath, hep-ph/0011177.

\bibitem{Kawamura:2000ir}
Y.~Kawamura,
hep-ph/0012352.

\bibitem{Chanowitz:1988ae}
M.~S.~Chanowitz,
Ann.\ Rev.\ Nucl.\ Part.\ Sci.\ {\bf 38}, 323 (1988).

\bibitem{cheng}
Hsin-Chia Cheng, hep-ph/0012263.

\bibitem{Coleman}
Coleman, S. Weinberg, E.\ Phys.\ Rev.\ {\bf D7}, 1888(1973).

\bibitem{Lahanas:1987uc}
A.~B.~Lahanas and D.~V.~Nanopoulos,
Phys.\ Rept.\ {\bf 145}, 1 (1987).

\bibitem{Antoniadis}
I.~Antoniadis,
Phys.\ Lett.\ B {\bf 246}, 377 (1990);
I.~Antoniadis and K.~Benakli,
Phys.\ Lett.\ B {\bf 326}, 69 (1994)
[hep-th/9310151].

\bibitem{Dienes:1999vg}
K.~R.~Dienes, E.~Dudas and T.~Gherghetta, Nucl.\ Phys.\  {\bf B537}, 47 (1999), [hep-ph/9806292].

\bibitem{Pomarol:1998sd}
A.~Pomarol and M.~Quiros,
Phys.\ Lett.\ {\bf B438}, 255 (1998) [hep-ph/9806263].

\bibitem{bando:1993}
M.~Bando, T.~Kugo, N.~Maekawa and H.~Nakano,
Prog.\ Theor.\ Phys.\ {\bf 90}, 405 (1993)
[hep-ph/9210229].

\bibitem{Casas:1999cf}
J.~A.~Casas, V.~Di Clemente and M.~Quiros,
Nucl.\ Phys.\ {\bf B553}, 511 (1999)
[hep-ph/9809275].

\bibitem{georgi}
H. Georgi, Annu. Rev.\ Nucl.\ Part.\ Sci.\ {\bf 43}, 209 (1993).

\bibitem{appelquist}
T. Appelquist and J. Carazzone, Phys.\ Rev.\ {\bf D11}, 2856 (1975).

\bibitem{hambye}
T. Hambye and K. Riesselmann, Phys.\ Rev.\ {\bf D55}, 7255 (1997).

\end{references}
\end{document}